\documentstyle[twocolumn,aps,epsfig]{revtex}

\begin{document}

\twocolumn[

\hsize\textwidth\columnwidth\hsize\csname@twocolumnfalse\endcsname

\draft

\preprint{conmatt}

\title{Atomic Scale Imaging and Spectroscopy of a CuO$_2$ Plane at the Surface of Bi$_{2}$Sr$_{2}$CaCu$_{2}$O$_{8+\delta}$}

\author{S. Misra, S. Oh, D.J. Hornbaker, T. DiLuccio, J.N. Eckstein, and A. Yazdani$^{*}$}

\address{Department of Physics and Fredrick Seitz Materials Research Laboratory, University of Illinois at Urbana-Champaign, Urbana, Illinois 61801}

\date{\bf{To appear in Physical Review Letters August 19}}

\maketitle

\begin{abstract}

We have used a scanning tunneling microscope to demonstrate that a single CuO$_2$ plane can form a stable and atomically ordered layer at the surface of Bi$_{2}$Sr$_{2}$CaCu$_{2}$O$_{8+\delta}$. In contrast to previous studies on high-T$_c$ surfaces, the CuO$_2$-terminated surface exhibits a strongly suppressed tunneling conductance at low voltages. We consider a number of different explanations for this phenomena and propose that it may be caused by how the orbital symmetry of the CuO$_2$ plane's electronic states affects the tunneling process. 

\end{abstract}


]


High-temperature cuprate superconductors are layered compounds containing one or more CuO$_2$ planes and other layers that act as charge reservoirs for these planes. It is believed that the strong electronic interactions in the CuO$_2$ planes are responsible for the cuprates' unusual electronic properties, including unconventional superconductivity with a $d$-wave order parameter and the non-Fermi liquid behavior found at low carrier concentrations \cite{Maple}. Consistent with this view, recent efforts have demonstrated that the phenomena observed in the bulk cuprates can be induced in a single CuO$_2$ plane using charge injection devices fabricated at the surface of the insulating infinite-layer compound CaCuO$_2$ \cite{Schon}. Therefore, it is both timely and of fundamental importance to study the electronic properties of a single CuO$_2$ plane at the surface. Surfaces of most materials from simple metals to compound semiconductors exhibit a variety of novel electronic phenomena; however systematic studies of layered cuprate surfaces have been limited. In particular, a study of surfaces terminated with different layers has not been accomplished.

Here we report the first imaging and electron spectroscopy measurements of a single CuO$_2$ plane at the surface of a cuprate superconductor using a scanning tunneling microscope (STM). Similar to most previous experiments that have probed the surfaces of cuprates \cite{Ginsberg,Hasegawa,Murakami,Kitazawa,Pan0,Sugita,Renner1,Renner2,Yazdani,Hudson,Pan1,Olson,Valla,Campuzano,Damascelli}, we carry out our studies on Bi$_{2}$Sr$_{2}$CaCu$_{2}$O$_{8+\delta}$. However, previous efforts have focused almost exclusively on surfaces terminated with the BiO layer, 4.5\AA\ above the pair of CuO$_2$ planes in this compound. Although there have been some reports of STM experiments on surfaces terminated with layers other than BiO, these efforts have failed to definitively identify the CuO$_2$ plane as the surface layer \cite{Hasegawa,Murakami,Kitazawa,Pan0,Sugita}. Such an identification must rely on a successful correlation of STM topographic images with crystallographic data, which depends, for example, both on preparing the samples' surface and on making measurements under ultra-high vacuum (UHV) conditions \cite{Kitazawa}.

Our measurements demonstrate that a CuO$_2$ plane in Bi$_{2}$Sr$_{2}$CaCu$_{2}$O$_{8+\delta}$ can form a stable and well-ordered surface layer with a lattice similar to that of the same plane in the bulk. In contrast to typical spectroscopic measurements on BiO-terminated surfaces, which exhibit an inchoate gap \cite{Renner2,Campuzano,Damascelli}, we find that tunneling into a CuO$_2$ plane at the surface is strongly suppressed at energies close to the Fermi level. This behavior, which is unexpected from a $d$-wave superconductor, raises the fundamental question of how a single CuO$_2$ plane behaves when it is at the surface and asymmetrically doped by layers beneath the surface. We consider a number of possible explanations for this phenomena including a model based on the effects of CuO$_2$ plane's orbitals on the tunneling process, which can produce suppression of tunneling at low energies.

\begin{figure} 

\begin{center}

\epsfig{file=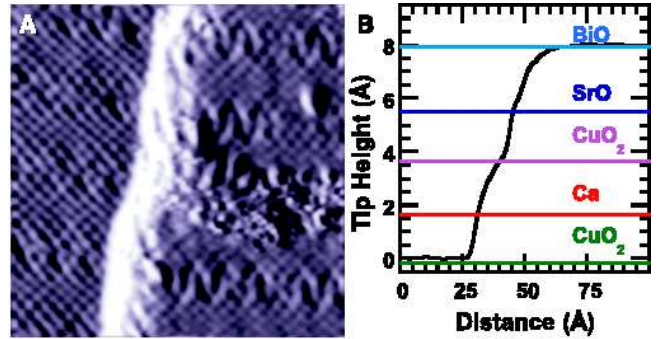,width=3.375in}

\caption{(A) A 100\AA$\times$100\AA\ STM topograph of a step edge, rendered to display the atomic corrugation on both the top (right) and bottom (left) of the step (V=-200mV and I=50pA). (B) Average of the topographic line scans in (A) and the known relative distances between the BiO plane and other crystallographic planes.}

\label{autonum}

\end{center}

\end{figure}

We performed our measurements using a home-built UHV STM operating at 4.2K on Bi$_{2}$Sr$_{2}$CaCu$_{2}$O$_{8+\delta}$ thin films (1000\AA) which were grown using molecular beam epitaxy (MBE). Resistivity measurements show the samples to have a T$_c$ of 84K and a transition width of 4K. The samples were introduced into the UHV chamber and cleaved mechanically at room temperature prior to making STM measurements. The majority of the areas of the nearly optimally doped samples show STM data similar to those previously obtained on the BiO-terminated surface of Bi$_{2}$Sr$_{2}$CaCu$_{2}$O$_{8+\delta}$ \cite{Renner1,Renner2,Yazdani,Hudson,Pan1}. This surface is typically exposed as a result of cleaving, a process in which the Bi$_{2}$Sr$_{2}$CaCu$_{2}$O$_{8+\delta}$ samples break at the weakest bond in the crystal structure, namely between the BiO planes of two adjacent half-unit cells \cite{Lindberg}. However, in addition to the BiO plane, we occasionally find smaller terraces terminated with crystallographic layers other than the BiO plane.

\begin{figure} 

\begin{center}

\epsfig{file=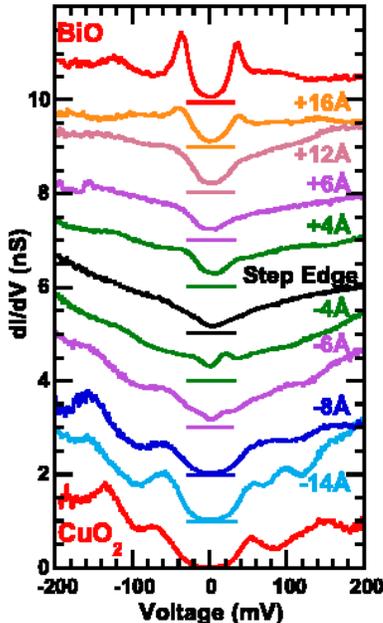,width=2.0in}

\caption{The STM spectra measured at the indicated distance perpendicular to the step edge shown in Figure 1a, offset for clarity. The topmost and bottom-most spectra are taken on the BiO and Cu0$_2$ planes respectively far away from the step edge.}

\label{autonum}

\end{center}

\end{figure}

We focus on situations such as that shown in Fig. 1, where we find a BiO terrace ending in a sub-unit cell height step. In Fig. 2, we show the spectra taken along a line perpendicular to the step edge in Fig. 1a. Tunneling spectra measured on the top terrace (on the right, Fig. 1a) are characteristic of those measured throughout the majority of the surface (Fig. 3c), and hence the top terrace is identified as the BiO plane. The spectra measured on the bottom terrace (on the left, Fig. 1a) are remarkably different from those measured on the top terrace, even when taking into account the extremes in variation of the BiO spectra (Fig. 3c). To identify the atomic terrace on the left side of Fig. 1a, we compare the average of the topographic line scans measured perpendicular to the step edge with the known separation between the different crystal planes in Fig. 1b \cite{Heinrich}. We have estimated various sources of inaccuracy in relating STM topography to structural information, such as those due to piezo calibration or to the difference in the electronic structure of the two planes. All such factors are far smaller (0.1-0.2\AA) than the expected distances between adjacent layers in the crystal structure ($\sim$2\AA). Hence, we can correlate the STM data and plane spacing as shown in Fig. 1b and conclude that the terrace on the left is the lower CuO$_2$ plane of Bi$_{2}$Sr$_{2}$CaCu$_{2}$O$_{8+\delta}$. Using this procedure, we have identified the CuO$_2$ plane as the terminating layer of distinct regions of several different samples grown under the same conditions. The coupling between CuO$_2$ planes in adjacent half-unit cells along the crystallographic c-axis in Bi$_{2}$Sr$_{2}$CaCu$_{2}$O$_{8+\delta}$ is extremely weak. Our measurements thus represent an examination of the electronic structure of a single CuO$_2$ plane that is asymmetrically doped via the layers beneath the surface.

Having identified a surface terminated with a CuO$_2$ plane, we
examine its electronic and structural properties far from step edges.
The high-resolution images of this surface in Fig. 3b show a
well-ordered lattice with an atomic spacing consistent with that
expected for a CuO$_2$ plane in the bulk crystal structure. The atomic
lattice also shows a periodic distortion due to the incommensurate
b-axis supermodulation known to run throughout the bulk of the
Bi$_{2}$Sr$_{2}$CaCu$_{2}$O$_{8+\delta}$  crystal. \cite{Heinrich}
Spectroscopic measurements (Fig. 3d) made on the surface CuO$_2$ plane
show a wider energy gap as compared with those typically measured on
the BiO surface, as well as a strong suppression of the tunneling
conductance within 10 meV of the Fermi level E$_F$. The unusual
tunneling characteristics measured on this plane are found to be
homogeneous on length scales from interatomic distances to several
hundreds of angstroms (gap value 60$\pm$10 meV). The strong contrast
between the spectra taken on the BiO and CuO$_2$ terminated surfaces
is the main result reported in this paper, which demonstrates the
dramatic influence of the layered structure on their surface electronic properties.

Surface sensitive measurements on the BiO-terminated surfaces is commonly interpreted to reflect the density of state (DOS) of the ''bulk" CuO$_2$ layers beneath surface. However, in the case of tunneling measurements performed on the BiO plane, the large separation between the tip and the pair of CuO$_2$ planes (greater than 10\AA) precludes the possibility of tunneling directly to these planes.  In this situation, the electronic states of planes other than CuO$_2$ planes play a role in the tunneling process and must be taken into account. Despite the lack of a clear model for this tunneling process, experimental support for interpretation of BiO spectra in terms of the DOS of the sub-surface CuO$_2$ layers has come from strong resemblance of STM spectra with those obtained by angle-resolved photoemission (ARPES) at $\vec{k}=(0,\pi)$ on the same surface. \cite{Ding} However, in this region of momentum space, the DOS of Bi$_{2}$Sr$_{2}$CaCu$_{2}$O$_{8+\delta}$ could be remarkably different from that expected for a single CuO$_2$ plane. Electronic structure calculations for Bi$_{2}$Sr$_{2}$CaCu$_{2}$O$_{8+\delta}$ have long predicted the presence of BiO-derived bands and a splitting of CuO$_2$ bands due to bilayer interactions at the Fermi level near $\vec{k}=(0,\pi)$. \cite{Singh} These two effects may have to be included to interpret both the STM and the ARPES data on BiO-terminated surfaces. To date, the contribution of the BiO layer to the DOS near E$_F$ has not yet been observed; however, various ARPES experiments have recently accumulated strong evidence for the predicted bilayer splitting of the CuO$_2$ bands. \cite{Damascelli} In general, the exact nature of ARPES measurements of the band structure near $(0,\pi)$ continues to be a subject of intense debate. \cite{Campuzano,Damascelli} The resolution of this debate is critical to the interpretation of tunneling spectra on the BiO surface and how they relate to our spectroscopic measurements of a single CuO$_2$ plane at the surface.

In contrast to tunneling on the BiO plane, direct tunneling into the surface CuO$_2$ plane does not involve the electronic structure of layers other than a single CuO$_2$. Despite this simplification, to interpret our measurements we must consider the possibility that this surface layer may have a dramatically different electronic character than either the same plane in the bulk or those underneath the BiO-terminated surface. Such a situation may arise as a result of structural modifications or changes in the doping level when the CuO$_2$ plane is at the surface. The STM images of the surface-terminated CuO$_2$ plane rule out the possibility of a structural reconstruction, as they show this plane has the atomic structure expected for a bulk CuO$_2$ plane in Bi$_{2}$Sr$_{2}$CaCu$_{2}$O$_{8+\delta}$. However, we can not rule out the possibility that the doping level is different at the surface and hence consider various electronic states that can give rise to the measured spectra.

One possibility would be that the surface CuO$_2$ plane is an insulator and the observed energy gap is associated with making single electron excitation in such a system. While we can not rule out this possible explanation, based on the measurements presented here we can place stringent constraints on possible insulating phases. First, the insulator must be electronically homogenous and have an energy gap of about 60 meV, a value unusually close to previously reported values of energy gap for highly underdoped (0.1 holes per Cu-O plaquette) but still superconducting samples.\cite{Tallon} Second, to explain our data, the insulator must have a particle-hole symmetric electronic structure with the chemical potential exactly in between the occupied and unoccupied bands. For bulk cuprates, at very low doping, there is an antiferromagnetic insulating ground state that has a 700 meV Mott wide gap \cite{Damascelli}, which is inconsistent with our measurements reported here. In addition to this phase, there have been proposals for a variety of interesting insulating bulk phases, such as spin or charge density waves states at low doping\cite{Sachdev} Many of these models not only involve inhomogeneous electronic states but also are expected to give rise to an inchoate pseudogap.

An alternative possibility would be that the surface CuO$_2$ plane is conducting, but the measured gap is due to Coulomb blockade, such as those observed when tunneling into disordered or granular thin films. Such a possibility can be diagnosed, as the size of the charging gap should be a sensitive function of capacitance.\cite{Coulomb} Our observation that the spectra is independent of tip-sample separation rules out charging effects as a potential explanation.

\begin{figure} 

\begin{center}

\epsfig{file=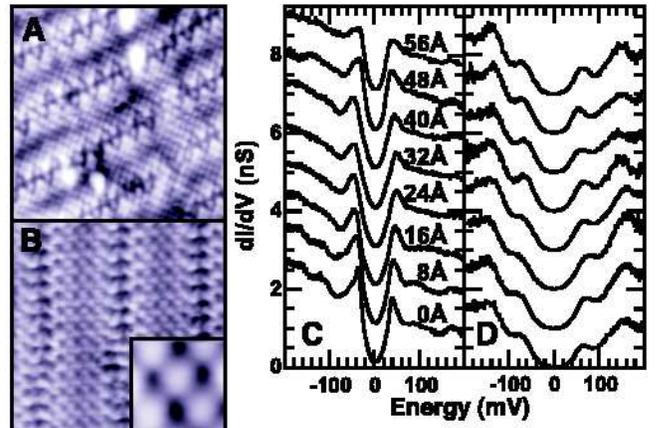,width=3.375in}

\caption{(A) A 100\AA$\times$100\AA\ STM topograph of the BiO plane. (B) STM topographs of the CuO$_2$ terminated surface. Main image (64\AA$\times$64\AA) and the inset (8\AA$\times$8\AA). (C) STM spectra of the BiO plane taken at the indicated distance along a straight line (offset for clarity). (D) STM conductance spectra taken along a straight line, at the indicated distances, on the CuO$_2$ plane (offset for clarity). All data taken with V=200mV and I=200pA.}

\label{autonum}

\end{center}

\end{figure}

Finally, we consider the possibility that the surface CuO$_2$ plane is in the superconducting state at low temperatures, although with a doping level that may be different than that of the bulk of our sample. Assuming a similar superconducting state as in the bulk, we would expect the density of states (DOS) of the surface layer at low energies to have a form $\rho_{s}(E,\vec{k})=\rho_{0}E/\sqrt{E^{2}-\Delta^{2}(\vec{k})}$, with an energy gap $\Delta(\vec{k})=\Delta_{0}[ cos(k_{x}a)-cos(k_{y}a)]$  that has the symmetry of a $d$-wave order parameter (a is the lattice constant). If we assume that the STM tunneling averages the DOS equally around the Fermi surface, the differential conductance $dI/dV$ is expected to be a linear function of energy near the E$_F$--a behavior that is in strong contrast to that measured on the CuO$_2$ plane. However, there have been theoretical efforts which suggest tunneling into a CuO$_2$ plane along the c-axis may not be averaging equally over the entire Fermi surface. \cite{Chakravarty,Anderson,Martin}

These previous efforts have shown that tunneling in the perpendicular direction into a CuO$_2$ plane is facilitated by the unoccupied 4s orbital of the Cu atoms, which extends further out-of-plane than the other orbitals in this plane.\cite{Chakravarty,Anderson,Martin}  With the STM tip perpendicular to the CuO$_2$ plane, we would thus expect the tip's electronic states to overlap most strongly with the Cu 4s orbitals and probe $\rho_{s}(E,\vec{k})$ near E$_F$ through these orbitals (Fig. 4a). However, the CuO$_2$ plane states near E$_F$ originate primarily from the Cu 3d band, which because of its d-orbital symmetry, couples to the Cu 4s orbitals with the matrix element $|M(\vec{k})|^{2}\propto [cos(k_{x}a)-cos(k_{y}a)]^{2}$ (Fig. 4a). \cite{Chakravarty,Anderson,Martin} Including the consequences of the orbital symmetry of the CuO$_2$ plane states on the tunneling process, we can relate the tunneling conductance to the DOS of a single CuO$_2$ plane as $dI/dV(V)\propto\int\int |M(\vec{k})|^{2}\rho_{s}(E_{F}+V,\vec{k})e^{-(E_{F}+V-U)^{2}/2\sigma^{2}}d\vec{k}dU$, 

where we have introduced Gaussian broadening (as characterized by $\sigma$) to account for energy broadening due to various sources. The effect of the matrix element is to filter the measurement of $\rho_{s}(E,\vec{k})$ preferentially for $\vec{k}$ directions in which it has a larger energy gap.

Figure 4b shows a comparison of the above model with the experimental data, using parameters to best match the tunneling conductance at low voltages, close to E$_{F}$.  The best fit to the experimental data is with $\Delta_{0}=60meV$, which is larger than that expected for the slightly underdoped samples (T$_c$=84K) used in our experiments. It is, however, close to previously reported values of the energy gap for highly underdoped (0.1 holes per Cu-O plaquette) but still superconducting, samples from various measurements. \cite{Tallon} The underdoped behavior of a single CuO$_2$ plane reported here may be the consequence of this layer being at the surface, where doping is provided to it only from below the surface--a situation different from that of a bulk CuO$_2$ plane. The model described above has the obvious shortcoming in that it does not address features at higher energies in the data. However, it does provide a simple physical argument, based on the orbital symmetry of the electronic states of a CuO$_2$ plane, for the apparent suppression of tunneling conductance at low biases.

\begin{figure} 

\begin{center}

\epsfig{file=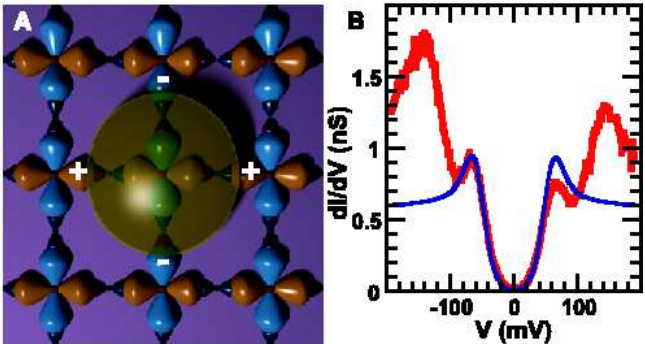,width=3.375in}

\caption{(A)The relevant orbital in the CuO$_2$ plane: O 2p (blue and purple),and Cu 3d (brown and blue), Cu 4s (yellow). A model of tunneling matrix elements based on these orbitals is introduced in the text. (B)Comparison of the model with tunneling spectra of the CuO$_2$ plane (red) for parameter values $\Delta$=60mV and $\sigma\sim$12mV.}

\label{autonum}

\end{center}

\end{figure}

In conclusion, we report the first STM imaging and spectroscopy measurements of a single CuO$_2$ plane at the surface of a high-T$_c$ cuprate. We find this layer to be atomically ordered and to exhibit an unusual suppression of tunneling DOS at low energies.  We have considered a number of possible explanations and have proposed that the orbital symmetry of the electronic states in the CuO$_2$ may be influencing the tunneling process and be responsible for the observed behavior at the surface.

We acknowledge discussions with P.W. Anderson, A.V. Balatsky, S.L. Cooper, J.C. Davis, M.E. Flatt$\acute e$, E. Fradkin, L.H. Greene, M.V. Klein, R.B. Laughlin, A.J. Leggett, Z.-X. Shen, D. Pines, P.W. Phillips, and D.J. Van Harlingen. This work was supported under NSF-CAREER(DMR-98-75565), ONR(N000140110071), NSF-FRG(DMR-99-72087), and DOE through FSMRL(DEFG-02-96ER4539).

\end{document}